\documentclass{ws-ijmpe}
          \usepackage[super,compress]{cite}
\usepackage{graphicx}
\usepackage{epstopdf} 
\usepackage[utf8]{inputenc}
\usepackage{float}
\usepackage{array}
\usepackage{multirow} 

\begin{document}

          \markboth{Gábor Riczu and József Cseh}{Gross features of the spectrum of the $^{36}$Ar nucleus}

\title{Gross features of the spectrum of the $^{36}$Ar nucleus }

%\author{G. Riczu}
%\affiliation{Institute for Nuclear Research, Debrecen, Pf. 51, Hungary-4001}
%\author{J. Cseh}
%\affiliation{Institute for Nuclear Research, Debrecen, Pf. 51, Hungary-4001}

%%%%%%%%%%\date{\today}

%\author{
%Gábor Riczu and József Cseh \\
%Institute for Nuclear Research, Debrecen, Pf. 51, Hungary-4001} 

\author{Gábor Riczu}

\address{Institute for Nuclear Research,\\
Debrecen, Pf. 51, 4001, Hungary\\
riczugabor@atomki.hu}

\author{József Cseh}

\address{Institute for Nuclear Research,\\
Debrecen, Pf. 51, 4001, Hungary\\
cseh@atomki.hu}

\maketitle

%\begin{abstract}
%We investigate the spectrum of the $^{36}$Ar nucleus, including a large range of the energy and deformation, observed in different reactions. In particular, the ground-state region, the superdeformed band and the candidate hyperdeformed band are described in a uniform manner in terms of the multiconfigurational dynamical symmetry. The basic features of the spectrum are reproduced to a reasonable approximation, and the coexistence of different cluster configurations is discussed.
%\end{abstract}

\begin{abstract}
Samples of the spectrum of the $^{36}$Ar nucleus are known in different energy windows. In addition to the ground state region (GS), the superdeformed (SD) state is observed, too, and there is a good candidate for the hyperdeformed (HD) one, as well. They are populated in different reactions. We intend to describe the gross features of the spectra of different energies, deformations and reactions in a unified way. We apply the multiconfigurational dynamical symmetry (MUSY). The SU(3) quantum numbers of the shape isomers from previous studies pave the way for this description. The MUSY reproduces the gross features of the spectra to a reasonable approximation. The energy spectrum of the three valleys (GS, SD, HD) indicates that the multiconfigurational symmetry is valid to a good approximation, and different cluster configurations coexist in the shape isomers.
\end{abstract}


%\ccode{PACS numbers: 21.60.−n, 21.60.Cs, 21.60.Fw, 21.60.Gx, 27.30.+t}
\ccode{PACS numbers: 21.60.-n, 21.60.Cs, 21.60.Fw, 21.60.Gx, 27.30.+t}

%\begin{keyword}
%multiconfigurational dynamical symmetry, excitation spectrum, clusterization, shape isomers
%\end{keyword}

%%%%%%%%%%\pacs{21.60.Fw, 21.60.Cs, 27.30.+t}

%keywords{
%quarteting and clustering, excitation spectrum, multichannel dynamical symmetry
%}

%%%%%%%%%%\oddsidemargin = -24pt

\section{Introduction}

The $^{36}$Ar nucleus has several interesting features, therefore, it has attracted considerable experimental and theoretical investigations. In addition to its low-lying spectrum 
\cite{Nica,Nndc,Iaea}
many of its highly-excited states have been observed in
$^{32}$S$(\alpha,\gamma)$$^{36}$Ar process, as well as in resonance reactions of heavy ions 
$^{20}$Ne+$^{16}$O and
$^{24}$Mg+$^{12}$C.

Theoretical structure studies have been performed within different approaches. Especially remarkable is the alpha-cluster calculation in the Bloch-Brink model, which indicated many stable configurations
\cite{Rae}. 

The most remarkable aspect of the structure of $^{36}$Ar
is probably the presence of the shape isomers. Its superdeformed (SD) state was observed in multiple gamma-coincidence experiments in the  
$^{24}$Mg($^{20}$Ne,2$\alpha)$$^{36}$Ar
reaction
\cite{Svensson}. 
Several structure model reproduced this state in good agreement with each other and with the experimental data.

The hyperdeformed (HD) state was predicted first from the alpha-cluster model
\cite{Rae}
(as a local minimum of the energy surface). 
Then the possible binary cluster-configurations of this state was investigated in
\cite{cs04}
(together with the clusterization of the ground and superdeformed states).
This study sheds some light not only on the structure of the deformed states, but also on the possible reaction channels in which they could be populated.
The 
$^{20}$Ne+$^{16}$O and
$^{24}$Mg+$^{12}$C
fragmentations turned out to be the favored ones.
Following these theoretical predictions in
\cite{Sciani} 
it was reported that the newly observed
$^{24}$Mg+$^{12}$C
resonances together with the known
$^{20}$Ne+$^{16}$O
ones seem to form a rotational band with the moment of inertia very close to that of the predicted HD state from the alpha-cluster calculation.
Furthermore, an independent theoretical approach, based on the quadrupole-shape stability and self-consistency study of 
\cite{Cseh}
predicted also a very stable HD
%hyperdeformed 
state in good agreement with the result of the cluster model and the experimental observation.
All these findings suggest that the observed rotational-like band is a very good candidate for  the hyperdeformed state of $^{36}$Ar, which is, therefore, the first self-conjugate nucleus with experimentally observed (or indicated) super- and hyperdeformed states.
Of course, gamma-coincidence measurement would be highly desirable in order to decide uniquely on the existence of the HD state. 
  
The connection between the deformation, clusterization and shell structure was found recently in terms of a dynamical symmetry 
\cite{ischia},
called multiconfigurational dynamical symmetry (MUSY).
It is the common intersection of the shell, collective and cluster models for the multi major shell problem. (See below for more details.) As a consequence it provides us with the relations of the wavefunctions as well as with the relations of the spectra of different approaches. 
%It is not very well suited for the reproduction of the fine details, but 
It seems to be very effective in  accounting for the gross features, and in connecting spectra of different configurations in different energy windows.

In case of the $^{28}$Si nucleus e.g. the MUSY was able to describe in a unified way the low-lying well-known bands and the high-lying cluster spectrum
\cite{cseri}. 
Not only that the 
$^{12}$C+$^{16}$O
resonance spectrum was obtained from the Hamiltonian of the quartet model
\cite{Cseh2} 
 of the ground-state region, but it was obtained as a parameter-free prediction. 
This study revealed that the 
$^{12}$C+$^{16}$O
resonances form the spectrum of the second minimum of the potential energy surface, which is built on the (recently discovered) superdeformed state
\cite{sisd1,sisd2}. 
Similarly, the highly excited spectrum of the $0^+$ states of the 
$^{24}$Mg+$^{4}$He
configuration was predicted by the same Hamiltonian, in good agreement with the experimental results
\cite{adsley}. 
This symmetry also reveals that seemingly very different cluster and shell configurations can be identical due to the effect of the antisymmetrization. For the $^{28}$Si states it is shown in detail in
\cite{epj}. 

In this paper we apply the multiconfigurational dynamical symmetry for the description of the 
$^{36}$Ar spectrum.  The main question we address is if a simple Hamiltonian (with an analytical solution) can account for the energy spectrum of a set of states distributed in the first (ground-state), second (superdeformed) and third (hyperdeformed) valley of the energy surface. In addition, these states are observed as different configurations:
shell,
$^{32}$S+$^{4}$He,
$^{24}$Mg+$^{12}$C, and
$^{20}$Ne+$^{16}$O.

From the viewpoint of the study of the shape isomers this work completes the symmetry-based previous investigations of
\cite{cs04,Cseh}.
So far the shape isomers were predicted, and their possible clusterizations were studied, but their energies were not determined.
Most of the structure models give the shape isomers as local minima of the energy surface, but the method applied in 
\cite{Cseh}
is different. 
It investigates the stability and the self-consistency of the quadrupole shape, as a function of the deformation parameters
\cite{lepcsok}. 
Actually, the  stability and the self-consistency of the SU(3) symmetry is investigated, but  its quantum numbers uniquely define the quadrupole shape
\cite{qdr,qro}. 
It is an alternative method to the well-known energy-minimum calculation, and the 
analogous results  from a basically independent procedure makes the theoretical predictions even more reliable. Furthermore, providing us with the SU(3) quantum numbers of the shape isomers, this approach has a direct connection to the reaction channels, via the selection rule.
Our present work gives the energy of the shape isomers, and by doing so it makes the symmetry-governed studies complete. Furthermore, it provides us not only with the energies for the ground-bands of the different minima, like ground-state, superdeformed and hyperdeformed, but also for the excited bands, i.e. for the whole spectrum (of different configurations).

\section{Multiconfigurational dynamical symmetry}
The multiconfigurational dynamical symmetry 
\cite{ischia}
connects the shell, collective and cluster models of the multi-major-shell problem,
as mentioned above. It has an interesting mathematical structure and physical content.
It contains a simple dynamical symmetry, defined by a single algebra-chain in each 
(shell, quartet or cluster) configuration, and a further symmetry 
which connects the configurations to each other. 
For the shell or quartet configuration it is usually the algebra-chain of the Elliott model 
\cite{elliott1,elliott2}
 \begin{eqnarray}
  U(3) \ \ \ &\supset& SU(3) \supset \ SO(3)  \supset \ SO(2) \\
\nonumber
[n_1,n_2,n_3]&,& (\lambda, \mu) ,   K,\ \ \  L, \ \ \ \ \ \ \ \    \ M ,
\label{u3chain} 
\end{eqnarray}
where only the space symmetries of the states are indicated. U(3) is a subalgebra of U(N), where N is the number of single-particle orbitals in the major shell. The spin-isospin part is characterized by Wigner's U$^{ST}$(4) group
\cite{wigner}, 
and the antisymmetry of the total wavefunction is guaranteed by the adjoint representations of U(N) and U$^{ST}$(4)
\cite{wybourne}.
In case of the multi-major shell problem the principle of the antisymmetrization in terms of the related U(3) and U$^{ST}$(4) representations is the same, though technically it is more complicated
\cite{epj}.

For a binary cluster configuration the space symmetry of the states is defined by 
\begin{eqnarray}
 U_{C_1}(3)   \otimes   U_{C_2}(3)  \otimes \  U_R(3)  \supset 
 U_C(3)  \ \otimes  U_R(3) \supset       U(3)  \supset SU(3) \supset SO(3)
%U(3)  \supset SU(3) \supset SO(3),
% \supset SO(2) ,
\label{sacmdysy}
\end{eqnarray}
\\
where $C_1$ and $C_2$ refers to cluster no 1, and 2, and the internal structure of the clusters is accounted for by the Elliott model
\cite{sacm}.
$R$ stands for the relative motion, which is described by the vibron model with algebraic structure of U$_R$(4)
\cite{vibron}.
In this case, too, the spin-isospin sector is described by U$^{ST}$(4), and the antisymmetry requirement is satisfied
\cite{sacm}.

For the many-major-shell problem, a unified classification scheme is provided by the algebra-chain
%\begin{align}
%\begin{eqnarray}
% U_x(3)  &\otimes  \ \ U_y(3)  \ \  \supset   \ U(3)  \supset \ SU(3)  \supset SO(3) \\
% \vert  [n^x_1,n^x_2,n^x_3] &,  [n^y_1,n^y_2,n^y_3] , \rho ,  [n_1,n_2,n_3]  , (\lambda , \mu) , K , L  \rangle \ \
%\nonumber 
%\label{eq:u3u3}
%\end{align}
%\end{eqnarray}
\begin{equation}
 U_x(3)  \otimes  \ \ U_y(3)  \ \  \supset   \ U(3)  \supset \ SU(3)  \supset SO(3)
\label{eq:u3u3}
%\end{align}
\end{equation}
for the shell, collective and cluster models
\cite{ischia}.
The extension of the Elliott model for the major shell excitations is the symplectic model
\cite{sympl1,sympl2,sympl3}.
The symplectic model has a contracted version
\cite{contr1,contr2},
which is a multi-major-shell collective model, based on bosonic degrees of freedom.
For the shell and collective models $x$ stands for the
band-head (valence shell), for the cluster model it refers to the internal
cluster structure. $y$ indicates in each case the major shell excitations;
in the shell and collective model cases it takes place in steps of $2 \hbar \omega$,
connecting oscillator shells of the same parity,
while in the cluster case it is in steps of $1 \hbar \omega$, incorporating all the 
major shells. For the  cluster model it has only completely
symmetric (single-row Young-tableaux) irreducible representations (irreps): $[n,0,0]$, while in the case of the
shell and collective models  it can be  more general.
As it is indicated by this unified classification scheme,  the model space of the three models have a considerable overlap, but they are not identical. 

The MUSY is a composite symmetry in the sense that the simple dynamical symmetries of the different configurations are connected to each other by a further symmetry. (This logical structure resembles to that of the dynamical supersymmetry of nuclear structure, where the simple dynamical symmetries of the bosonic and fermionic sectors are connected by the supertransformations
\cite{susy1,susy2,susy3}.) 
In the MUSY case the connecting transformations are those of the pseudo-space of the particle indices
\cite{musy31,musy32}.
It is not visible in the shell-like scheme of chain 
(\ref{eq:u3u3}),
rather one should look at a classification scheme of the A-particle problem, in which the particle indices are explicitly included
\cite{musy31,musy32,kramosh}.
 
The multiconfigurational dynamical symmetry is obtained when one choses a Hamiltonian which is invariant with respect to the transformations in the pseudo-space of particle-indices. This is the case, if it is expressed in terms of the operators of the second part of chain
(\ref{eq:u3u3}). A particularly simple Hamiltonian of such kind is written in terms of the Casimir operators of the algebras
U(3) $\supset$ SU(3) $\supset$ SO(3).

Therefore, the two pillars of the multiconfigurational dynamical symmetry are 
i) the unified classification scheme 
(\ref{eq:u3u3})
of the shell (or quartet), and collective as well as cluster states, and
ii) a Hamiltonian, which is symmetric with respect to the transformations (in the pseudo-space of particle indices), that connect the different configurations. 

Due to the microscopic treatment of the model spaces, i.e. all the nucleon degrees of freedom are taken into account, and the Pauli-principle is appreciated, the different configurations may overlap with each other. An especially interesting case is when this overlap is 100\%. Such situations can easily be realized  once we construct the  full no-core shell model space up to a certain excitation number. The shell model basis is complete, therefore, any state vector can be expanded in this basis, and the basis states belonging to different SU(3) irreps are orthogonal to each other. Thus in case the multiplicity of the shell model basis of a specific SU(3) irrep is 1, then all the (different cluster) configurations are identical with it, having only a single term in the shell-model expansion. As a consequence, the total overlap of different configurations of basis 
(\ref{eq:u3u3})
is easy to be detected. Furthermore, the ground state, and the shape isomers usually have very good SU(3) symmetry
\cite{lepcsok,dubna},
so the symmetries of the basis states illuminate the fact that seemingly different configurations can be identical with each other, as a result of the antisymmetrization.

A further interesting feature of the MUSY is that it shows a dual breaking of symmetries
\cite{epj}. 
In this respect it  is similar to the dynamical symmetry of the Elliott model as well as to many other dynamical symmetries of the nuclear structure models. In particular, the U(3) and SU(3) symmetries are dynamically broken by the symmetry breaking interaction, represented by the invariant operator of the SO(3) algebra
\cite{spontan}. 
On the other hand the total Hamiltonian separates into an intrinsic (U(3) and SU(3) dependent), and a collective (SO(3) dependent) parts. In other words, the fast and slow degrees of freedom are separated. Both parts of the Hamiltonian are SO(3) invariant, but the ground state (and many other states) of the intrinsic Hamiltonian are not rotationally invariant. Thus the SO(3) symmetry is spontaneously broken in the eigenvalue problem of the intrinsic Hamiltonian. Furthermore, as pointed out in the previous paragraph, the nonspherical shape of the intrinsic state can have seemingly different configurations,
but the differences might be washed out by the antisymmetrization. When the total Hamiltonian is considered, then the (rotational) symmetry is recovered, as it is usual in the spontaneous breaking 
\cite{spontan}.

\section{Experimental data}
As described in the introduction, the GS and SD bands of the $^{36}$Ar are known experimentally.
(We have considered the 6$^{+}$ state at 9.182 MeV as a member of the ground-band,
in line with Refs.
\cite{Nndc,Svensson}.)
Furthermore, there is a promising candidate for the HD band \cite{Sciani}. 
In the compilation
\cite{Abbondanno}
one sees seven further $^{24}$Mg+$^{12}$C and two $^{20}$Ne+$^{16}$O resonances
in addition to those incorporated in
\cite{Sciani}. 
In Fig.1. we have shown also these states.
The moment of inertia of this band is 11.36 $\hbar^2/MeV$.

\begin{figure}[h]
\begin{center}
\includegraphics[height=7.0cm,angle=0.]{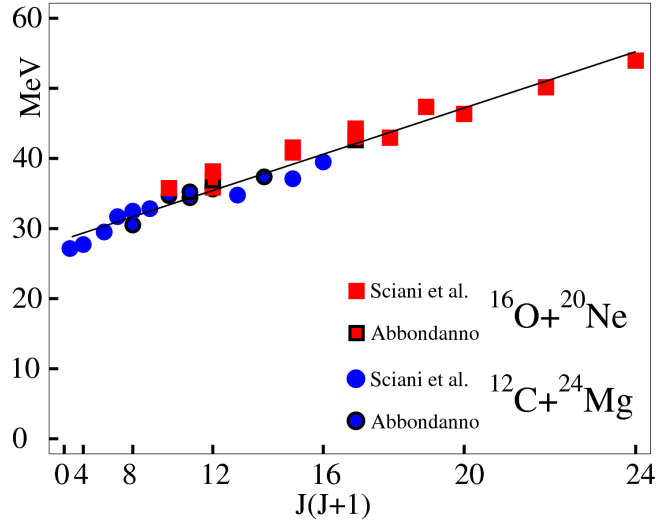}
\caption{$^{12}$C+$^{24}$Mg and $^{16}$O+$^{20}$Ne resonances in the $^{36}$Ar nucleus as a function of J(J+1).
The similar figure of Ref.
\cite{Sciani}) is extended by further resonances form
\cite{Abbondanno}.}
\end{center}
\end{figure}

In the low-lying region we arranged some states into bands. In particular, we did so, when the energies of more than two states
fall on the linear of J(J+1)
\cite{Nica,Nndc,Iaea}.
In some cases in-band E2 transitions are also known, as indicated in the lower part of Fig.2. Based on these criteria we guess one positive parity and two negative parity bands. 
We consider them merely as some loose indication, our main emphasis is the investigation of the ground-band and shape isomers in a unified framework.

\section{Calculation of the spectrum}
Here we discuss the application of the MUSY for the description of the $^{36}$Ar. For that, we assigned the model bands to the experimental ones. 
In doing so, we divided the HD band into parts of positive and negative parities (Fig. 2),
in line with our model calculation.

%\begin{figure}[H]
%\begin{center}
%\includegraphics[height=10.5cm,angle=0.]{36Arresmanu.eps}
%\caption{$^{24}Mg+^{12}C$ and $^{20}Ne+^{16}O$ resonances in the $^{36}Ar$ nucleus as a function of $J(J+1)$.}
%\end{center}
%\end{figure}

In Ref. 
\cite{Cseh} 
the U(3) quantum numbers of the 
GS, SD and HD states were determined from a symmetry stability and self-consistency calculation. Here we have applied the quantum numbers from
\cite{Cseh}.  
The other HD band (of negative parity) are associated with the band that was closest to the positive band in deformation and had appropriate spin-parity content.
Finally, we assigned the most deformed representations of 0 and 1 $\hbar\omega$ model space to the three low-energy bands, that had appropriate spin-parity content. 

We have applied a simple MUSY 
%U(3) dynamically symmetric 
Hamiltonian, which has an analytical solution in the
U(3) $\supset$ SU(3) $\supset$ SO(3)
basis
\cite{cseri}:
\begin{equation}
\hat{H}=(\hbar\omega)\hat{n}+a\hat{C}^{(2)}_{SU(3)}+b\hat{C}^{(3)}_{SU(3)}+d\frac{1}{2\theta}\hat{L}^2.
\end{equation}

The first term is the harmonic oscillator Hamiltonian (linear invariant of the U(3)), with a strength obtained from the systematics \cite{Blomqvist} $\hbar\omega$=45A$^{-\frac{1}{3}}$-25A$^{-\frac{2}{3}}$MeV=11.335 MeV. The second order invariant of the SU(3) (C$^{(2)}_{SU(3)}$) represents the quadrupole-quadrupole interaction. Its expectation value for an SU(3) basis state with quantum numbers $(\lambda,\mu)$ is $\lambda^2+\mu^2+\lambda \mu+3(\lambda+\mu)$.
The third order Casimir-operator (C$^{(3)}_{SU(3)}$) with eigenvalues $(\lambda-\mu)(\lambda+2\mu+3)(2\lambda+\mu+3)$
splits the degeneracy of
%distinguishes between 
the prolate $(\lambda > \mu)$ and oblate $(\lambda <  \mu)$ shapes with identical eigenvalues of the second order Casimir. $\theta$ is the moment of inertia calculated classically for the rigid shape determined by the $[n_1,n_2,n_3]$ U(3) quantum numbers \cite{Cseh3}. In particular, the ratio of the semi-major axes $z$, $x$, and $y$ is  obtained from a self-consistency argument \cite{Bohr}:
\begin{equation}
\frac{z}{y}=\frac{n_1+\frac{A}{2}}{n_3+\frac{A}{2}} ,\ \ \ \ \
\frac{x}{y}=\frac{n_2+\frac{A}{2}}{n_3+\frac{A}{2}} .  \\
\end{equation}
Their lengths are determined by the volume conservation:
\begin{equation}
y=R_0\sqrt[3]{A{\frac{(n_3 + \frac{A}{2})^2}{(n_1 + \frac{A}{2})(n_2 + \frac{A}{2})}}} \  .
\end{equation}
(Here we applied $R_0= 1.2 fm$.) The moment of inertia (in units of $\frac{\hbar^2}{MeV}$) for a rotor with axial symmetry ($x=y$) is given by
\begin{equation}
\theta=\frac{1}{5}m(z^2+x^2)  .
\end{equation}
%Please note that the energy eigenvalue problem has an analytical solution with this part, too. 
This Hamiltonian proved to be useful in a unified description of low-lying quartet spectrum and high-lying cluster spectrum in other examples, too
\cite{cseri, adsley}. 
So far, however, it was applied only for the spectra of the first (ground-state) and second (superdeformed) valley of the energy-surface. Here we check its applicability with respect to the simultaneous description of the spectra of three local minima. 

The parameters $a, b$ and $d$ were fitted to the experimental data:
 $a$=-0.11 MeV, $b$=0.00047 MeV, $d$=1.03.
  (In the fitting procedure the better-known GS and SD bands had a unit weight, the other bands had weight of 0.01.)
  
  The in-band B(E2) value is given as
\begin{eqnarray}
B(E2,L_i\rightarrow L_f)=\frac{2L_f+1}{2L_i+1}\alpha^2 %\nonumber \\
|\langle(\lambda\mu)KL_i,(11)2||(\lambda\mu)&KL_f\rangle|^2C^2_{SU(3)} ,
\end{eqnarray}
where $\langle(\lambda\mu)KL_i$,(11)2$||(\lambda\mu)KL_f\rangle$ is a SU(3) $\supset$ SO(3) 
Wigner coefficient \cite{Wig}, and $\alpha^2$ (=0.466 W.u.) is a parameter fitted to the experimental value of the 2$^+_1 \rightarrow$ 0$^+_1$ transition of 8.2 W.u.
  
\begin{figure}[H]
\begin{center}
\includegraphics[height=9.95cm,angle=0.]{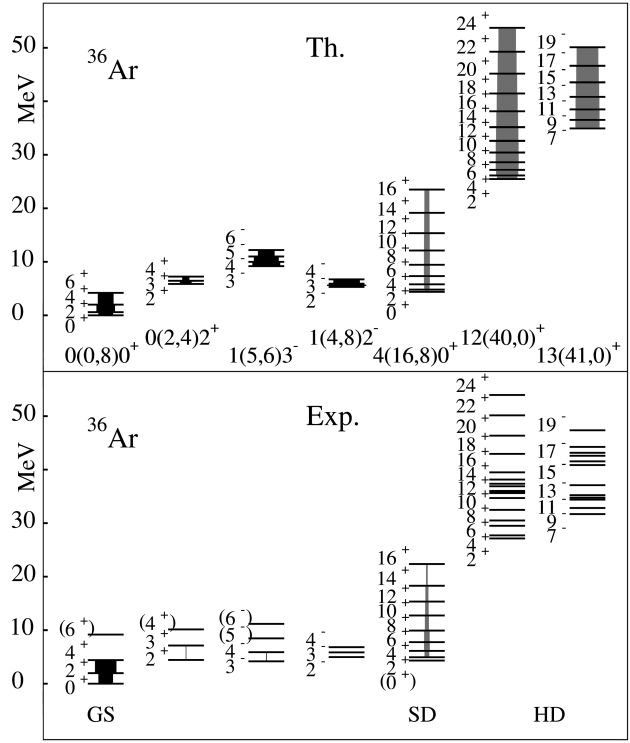}
\caption{The spectrum of the 
%semimicroscopic algebraic quartet model 
multiconfigurational dynamical symmetry (upper part)
in comparison with the experimental data of the  $^{36}$Ar nucleus (lower part). The experimental ground, super and hyperdeformed bands are labeled by GS, SD and HD, and the model bands by the n$(\lambda\mu)$K$^{\pi}$ labels. The width of the arrow between the states is proportional to the strength of the E2 transition. The real strength of the gray arrows (of the SD and HD bands) are 20 times of the illustrated ones.}
\end{center}
\end{figure}

%\begin{figure}[H]
%\begin{center}
%\includegraphics[height=10cm,angle=0.]{36Ardensity.eps}
%\caption{The spectrum of the semimicroscopic algebraic quartet model in comparison with the experimental data in the 5-12 MeV energy window in the  $^{36}Ar$ nucleus. The experimental spectra \cite{Nndc} are under the angular momenta on the left side, and the calculated spectra on the right side. States of uncertain spin-parity are indicated by red, and the certain states by black. The number of states with the appropriate colors are listed below the columns.}
%\end{center}
%\end{figure}

\begin{figure}[h]
\begin{center}
\includegraphics[height=4.5cm,angle=0.]{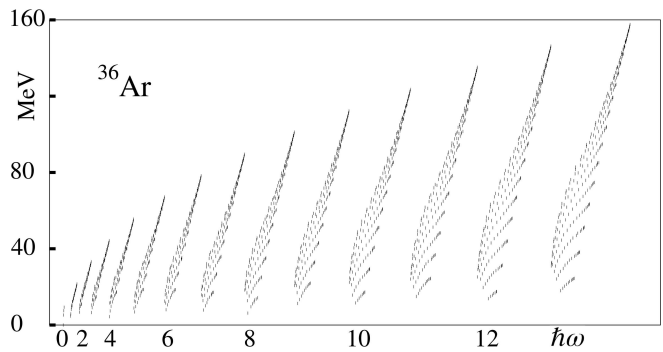}
\caption{The landscape of the quartet band-heads in the $^{36}$Ar nucleus.}
\end{center}
\end{figure}

%Fig. 4 shows the comparison between the experimental \cite{Nndc} and theoretical level densities in the 5-12 MeV energy window. The reason for choosing this energy window is that the 85\% of the experimental states are in this interval. All other states located rarely below and above this interval. States of uncertain spin-parity are indicated by red, and the certain states by black. (Note that the number of the calculated states is usually larger than the number of the illustrated, because there is no K dependence in the Hamiltonian.) Considering the uncertainty of the number of experimental states, and that this densities spectrum is calculated without fitting to the expermental level-density, the result is in good agreement with the experimental observations.

The distribution of the quartet (T=0, S=0) band-heads are shown in Fig.3. In particular, the lowest-lying state of each SU(3) $(\lambda, \mu)$ representation is plotted in the 0-13 $\hbar \omega$ major shells. They have either $0^+$, or $1^-$ spin-parity due to the relation of the Elliott quantum numbers of chain (1):
$K=min(\lambda, \mu), min(\lambda, \mu)-2,...,1$ or $0$;
$L=K, K+1, ..., K+max(\lambda,\mu)$, except for $K=0$, when $L=max(\lambda,\mu), max(\lambda,\mu)-2, ..., 1$ or 0. The energies are given by Eq. (4), with the parameters given above.

We determined the shapes of the investigated states (Fig. 4). From the shell-model side the quadrupole shape is given by the U(3) quantum numbers of the state \cite{Cseh3}. The  $^{20}$Ne+$^{16}$O, $^{24}$Mg+$^{12}$C, $^{32}$S+$\alpha$ cluster configurations can be obtained from the Harvey prescription \cite{Harvey1,Harvey2} and from the U(3) selection rule \cite{Selection1,Selection2,Selection3} which describes the structural aspect of the fusion (or fission) of a nucleus in terms of the harmonic oscillator basis. Since the multiplicity of the relevant U(3) representation is 1 in the shell basis, these shell and cluster configurations turn out to be identical with each other, due to the effect of the antisymmetrization. \\

\begin{figure}[h]
\begin{center}
\includegraphics[height=9.2cm,angle=0.]{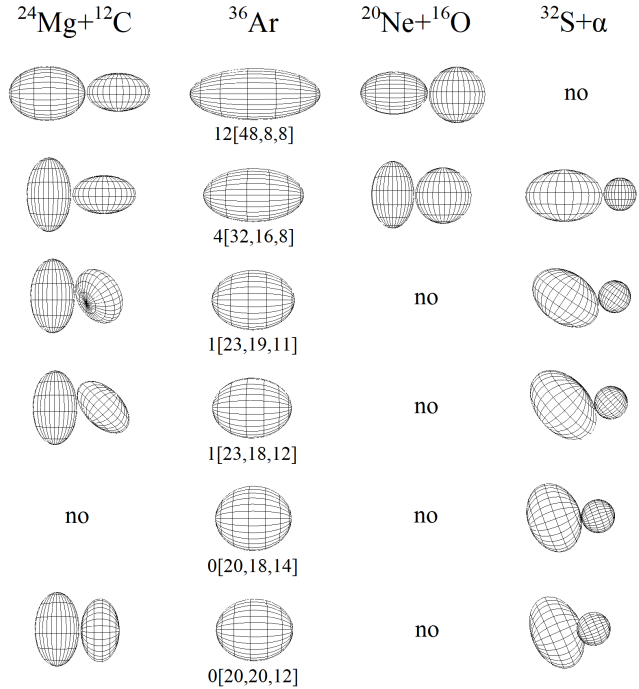}
\caption{Shape of some states in $^{36}$Ar in increasing energy order. In  [ ] parenthesis, the U(3) labels are indicated, while the first integer shows the major shell excitation quanta. Note, that the multiplicity of these U(3) states in the shell basis is 1, therefore, the indicated shell, and  cluster configurations have wavefunctions with 100\% overlap in each case, as a consequence of the
antisymmetrization.}
\end{center}
\end{figure}

Figure 5 shows the relationship between different clusterizations. Here the energies of band-heads of 0-4 $\hbar\omega$ cluster model spaces are presented.
Please, note that  increasing energy corresponds to decreasing eigenvalues of second-order SU(3) Casimir operator (within each major shell), i.e. decreasing deformation. It can be seen that the ground state and the most deformed states are generated by $^{32}$S+$\alpha$ and $^{24}$Mg+$^{12}$C clusterizations, while the less deformed states occur only in $^{32}$S+$\alpha$ configuration. Furthermore, the overlap of  $^{32}$S+$\alpha$ and $^{24}$Mg+$^{12}$C systems is significant, and the $^{20}$Ne+$^{16}$O configuration appears only from 4 $\hbar\omega$ excitation. 
%These statements are also true in major shells of  higher excitations.

\begin{figure}[h]
\begin{center}
\includegraphics[height=4.3cm,angle=0.]{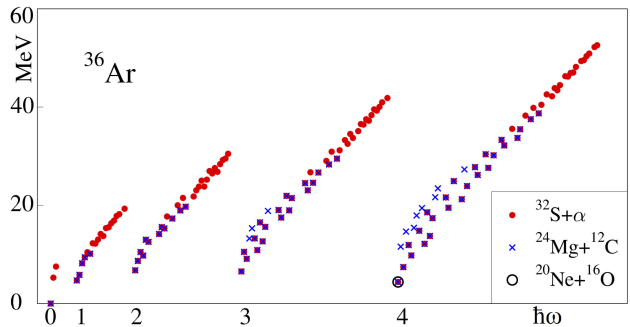}
\caption{Band-heads (L=0 or 1) of 0-4 $\hbar\omega$ cluster model spaces in the $^{36}$Ar.} 
%They are shown in decreasing order of the eigenvalues of $\hat{C}^{(2)}_{SU(3)}$ within each major shell.}
\end{center}
\end{figure}

\section{Summary and conclusions}

In this paper we have investigated the spectrum of the $^{36}$Ar nucleus, observed in different energy windows and in different reactions. This nucleus is special, inasmuch its superdeformed state is known from multiple coincidence  experiment
\cite{Svensson},
it has a good candidate for the hyperdeformed state
\cite{Sciani},
and it  allows several self-conjugate cluster configurations.

We have applied a symmetry-governed approach in studying the question whether or not the spectra of the first (ground-state), second (superdeformed) and third (hyperdeformed) valley can be described in a unified way. In particular, we have applied the multiconfigurational dynamical symmetry, which is the common intersection of the shell, collective and cluster models for the multi-major-shell problem. Due to this feature the MUSY can be able to account for low-lying shell-like and high-lying cluster spectra in a joint framework, incorporating also the deformation of the states. 

In previous symmetry-based studies of the $^{36}$Ar nucleus
the shape isomers have been determined 
\cite{Cseh}.
They were obtained from the investigation of the stability and the self-consistency of the U(3) symmetry, which is uniquely related to the the stability and self-consistency of the quadrupole deformation
\cite{lepcsok}.  This new method is an alternative to the well-known energy-minimum calculations. It provides us with the U(3) symmetry of the states, and therefore, allows the application of  a selection rule for the allowed cluster configurations. On the other hand it does not give the energy of the states. Our present study complete these investigations, by determining the energy of the states not only for the lowest-lying band in each valley, but for the detailed spectrum.
 %As a result a series of shape isomers were obtained, in good agreement with the results of the energy-minimum calculations, and in case of the  super-, and hyperdeformed state also with the experimental observation.
 
Here we have found that the gross features of the spectra of the three valleys  can simultaneously be accounted for by the MUSY. In the low-energy region we have considered the ground-band a further positive parity and two negative parity bands, which could be anticipated from the  experimental data. In the second valley the superdeformed band, while in third one the candidate hyperdeformed band was considered, split into a positive and a negative parity parts. It is remarkable that a simple Hamiltonian with analytical solution  reproduces the gross features of the observed spectrum. 

This finding resembles the example of the 
%indicates that the multiconfigurational dynamical symmetry is a good approximation to the spectrum of the $^{36}$Ar nucleus. In the case of the 
$^{28}$Si nucleus. In that case it was possible even to  extrapolate the high-lying cluster spectra from the low-lying quartet spectrum
\cite{cseri}. 
There a rich spectrum of well-defined low-lying bands was available, including also the SU(3) symmetry of the states (prior to our study).
In case of the $^{36}$Ar such an extrapolation does not work. Here much less information is available on the low-lying spectrum, so there is no real basis for an extrapolation. Therefore, in the fitting procedure we have taken into account the shape isomers, too. Nevertheless, it is remarkable, that a very small (0.01) weight of the HD states was enough to obtain a good fit. 
The multiconfigurational symmetry turns out to be approximately  valid,
what is remarkable, especially in light of the differences of the configurations, energy windows and deformation parameters.

We have studied the similarities of different cluster configurations, too. The 
$^{32}$S+$^{4}$He,
$^{24}$Mg+$^{12}$C, and
$^{20}$Ne+$^{16}$O, fragmentations
are known to be relevant for the $^{36}$Ar,
in the sense that these reaction channels have been studied experimentally. (In these configurations the nuclei are supposed to be in their ground intrinsic state, i.e. no nucleon excitations are taken into account, but collective rotations are possible.)   
It turned out that in several states, like e.g. ground-, SD, and HD states some cluster configurations and the shell configuration has complete overlap, 
%(within the dynamical symmetry approximation). 
as a consequence of the antisymmetrization (see Figure 4). This statement is valid to the extent the U(3) symmetry is a good for the wavefunctions, but it is known to be especially good for the shape isomers
\cite{lepcsok}. 
(See also the symmetry-diagnostics of the most general shell model wavefunctions of light nuclei
\cite{johnson1,johnson2}.)
The distribution of these cluster configurations along the energy scale is illustrated in Figure 5, indicating a considerable overlap between the 
$^{32}$S+$^{4}$He,
$^{24}$Mg+$^{12}$C, 
clusterizations, which are dominant in the low-lying region.
The 
$^{20}$Ne+$^{16}$O, 
configuration starts to play an important role around and above the superdeformed state.

\section*{Acknowledgements}
This work was supported by the
National Research, Development and Innovation Fund of Hungary, 
financed under the K18 funding scheme with project no. K 128729. We also acknowledge
KIFÜ for awarding us access to resource based in Hungary at Debrecen/Szeged.
Fruitful discussion with G. L\'evai is kindly acknowledged.

%%%%%%%%%%\noindent
%%%%%%%%%%{\it This work was supported by the
%%%%%%%%%%National Research, Development and Innovation Fund of Hungary, 
%%%%%%%%%%financed under the K18 funding scheme with project no. K 128729. We also acknowledge
%%%%%%%%%%KIFĂ for awarding us access to resource based in Hungary at Debrecen/Szeged.
%%%%%%%%%%Fruitful discussion with G. L\'evai is kindly acknowledged.
%%%%%%%%%%}

\end{document}